\documentclass[compsoc, conference, a4paper, 10pt, times]{IEEEtran}

\usepackage{booktabs}
\usepackage{cite}
\usepackage{amsmath,amssymb,amsfonts}
\usepackage{algorithmic}
\usepackage{graphicx}
\usepackage{textcomp}
\usepackage{bmpsize}
\usepackage[table]{xcolor}
\usepackage{lipsum}
\usepackage[colorlinks=true,urlcolor=black,breaklinks]{hyperref}
\usepackage{subfig}
\usepackage{todonotes}
\usepackage{multirow}
\usepackage{enumitem}
\usepackage{makecell}
\usepackage{url}
\usepackage{textcomp}
\newcommand{\etal}{\emph{\mbox{et al.~}}}

\newcommand{\comment}[1]{}
\makeatletter
\newcommand{\linebreakand}{%
  \end{@IEEEauthorhalign}
  \hfill\mbox{}\par
  \mbox{}\hfill\begin{@IEEEauthorhalign}
}
\makeatother
\usepackage{ragged2e}
\newcolumntype{L}[1]{>{\RaggedRight}p{#1}}

\begin{document}

\title{Assessing Network Operator Actions to Enhance Digital Sovereignty and Strengthen Network Resilience: A Longitudinal Analysis during the Russia-Ukraine Conflict}

\author{
\IEEEauthorblockN{Muhammad Yasir Muzayan Haq}
\IEEEauthorblockA{
\textit{University of Twente}\\
Enschede, The Netherlands \\
m.y.m.haq@utwente.nl}
\and
\IEEEauthorblockN{Abhishta Abhishta}
\IEEEauthorblockA{
\textit{University of Twente}\\
Enschede, The Netherlands\\
s.abhishta@utwente.nl}
\and
\IEEEauthorblockN{Raffaele Sommese}
\IEEEauthorblockA{
\textit{University of Twente}\\
Enschede, The Netherlands\\
r.sommese@utwente.nl}
\linebreakand
\IEEEauthorblockN{Mattijs Jonker}
\IEEEauthorblockA{
\textit{University of Twente}\\
Enschede, The Netherlands \\
m.jonker@utwente.nl}
\and
\IEEEauthorblockN{Lambert J.M. Nieuwenhuis}
\IEEEauthorblockA{
\textit{University of Twente}\\
Enschede, The Netherlands\\
l.j.m.nieuwenhuis@utwente.nl}
}

\maketitle

\begin{abstract}
We conduct longitudinal and temporal analyses on active DNS measurement data to investigate how the Russia-Ukraine conflict impacted the network infrastructures supporting domain names under ICANN's CZDS new gTLDs. Our findings revealed changes in the physical locations of network infrastructures, utilization of managed DNS services, infrastructure redundancy, and distribution, which started right after the first reported Russian military movements in February 2022. We also found that domains from different countries had varying location preferences when moving their hosting infrastructure. These observed changes suggest that network operators took proactive measures in anticipation of an armed conflict to promote resilience and protect the sovereignty of their networks in response to the conflict.
\end{abstract}



\section{Introduction}

In today's world, armed conflicts between countries are often supplemented by cyber attacks and attempts at information snooping \cite{lin2012cyber}. The recent invasion of Russia on Ukraine resulted in the destruction of several cities within Ukraine. Although the armed assault on Ukraine began on February 24th, 2022, the first signs of military build-up were reported nearly one year prior. During this conflict, Russian forces also targeted the network infrastructure within Ukraine. According to reports, since the beginning of the invasion, more than 4,000 base stations and 60,000 kilometers of fiber-optic lines used for the Internet belonging to Ukrainian telecommunications providers were seized or destroyed by Russian forces \cite{bergengruen_battle_2022}.

The resilience and robustness of network infrastructure are crucial for the seamless functioning of our digital society. A resilient Internet maintains an acceptable level of service despite faults and challenges to normal operation \cite{noauthor_internet_nodate}. Network operators optimize their service levels by improving physical network infrastructure, avoiding single points of failure, and managing network services. However, natural disasters such as volcanic eruptions, earthquakes, and armed conflicts between nations can lead to the destruction of critical network infrastructure and severely hamper the services of network service providers located in the affected region.

The threat of physical damage and cyber attacks, such as denial-of-service attacks, to hosting infrastructure is a pressing concern for network managers. To ensure the resilience of network services, they might take preemptive measures. During military conflicts between countries, interception of enemy communications is a common tactic used to compromise digital sovereignty. To mitigate this threat, network providers are incentivized to migrate their network services out of the enemy's physical territory or reduce dependence on infrastructure in conflict areas. One option is to make use of managed hosting or managed DNS service providers who offer more robust and resilient services for a premium \cite{50ab3fadbc7341e9bf8a5c757b757fd1}. 

This paper presents an analysis of the steps taken by network operators to enhance the resilience of network infrastructure and preserve digital sovereignty during the Russia-Ukraine conflict, based on large-scale Internet measurement data. 
\section{Timeline of the Conflict}

The conflict between Russia and Ukraine started way back in 2014 with the annexation of Crimea by Russia followed by the Donbas War~\cite{noauthor_donbas_war_2023}. However, in this paper, we focus on the full-scale invasion which officially started on February 24th, 2022.

\vspace{0.5em}
\noindent\textbf{\textit{Pre-invasion tensions.}}
On February 21st, 2021, the Russian Defense Ministry announced the deployment of 3,000 paratroopers to the Russia-Ukraine border~\cite{noauthor_untold_2022}. The soldiers were said to be deployed for large-scale exercises. In March, further military personnel and equipment massed up. While troops were partially withdrawn by mid-2021, a second buildup started in October, and by the end of 2021 a hundred thousand soldiers were heaped up around Ukraine's borders with Russia and Belarus, as well as within Crimea~\cite{noauthor_prelude_2023}.

\vspace{0.5em}
\noindent\textbf{\textit{Start of invasion}}
A year after the previously discussed Russian Defense Ministry announcement, Russia officially recognized the Donetsk People's Republic and the Luhansk People's Republic, as independent states, and deployed troops to the Donbas region. One day later, on February 22nd, 2022, Russia formally withdrew from the Minsk protocol. Two days later, Russia launched a full-scale invasion of Ukraine~\cite{noauthor_putin_announcement_2023}. Russian forces began shelling and missile strikes, focusing on airfields, military depots, and centers of Ukraine's military administration.

\vspace{0.5em}
\noindent\textbf{\textit{Attacks on Ukrainian Infrastructures.}}
To date, Russia has performed indiscriminate attacks in densely populated areas and has ramped up deliberate, systematic attacks on civilians and civilian infrastructures~\cite{noauthor_russian_strikes_2023}. In October 2022, Russia caused colossal damage to Ukraine's critical energy infrastructure and launched airstrikes on residential buildings in a number of cities, reportedly in retaliation for the Crimean Bridge explosion~\cite{noauthor_russia_energy_attack_2022}. 
Network and data infrastructures also became the target of physical attacks, especially when cyberattacks failed\cite{noauthor_ukraine_nodate}. Therefore, since the start of the conflict, big tech companies have been helping the Ukrainian government move their sensitive data to the cloud\cite{sanger_as_2022,stupp_ukraine_nodate}.

\section{Related Works}
\label{sec:related_works}
\vspace{-0.2em}
\noindent\textbf{\textit{Impact of Russia-Ukraine conflict on Internet ecosystem.}}
Jonker \etal\cite{jonker_where_2022} investigate the impact of the conflict in Ukraine on Russian domain infrastructure by conducting an empirical research to measure changes in naming, hosting, and certificates. They show that effects of barriers created by the war scenario manifest both as internal pressures on Russian sites to (re-)patriate the infrastructure they depend on (e.g., naming and hosting) and external pressures arising from Western providers disassociating from some or all Russian customers. By analyzing the impact of conflict on Russian domain infrastructure, the authors shedded light on how internal and external pressures affect Russian sites' infrastructure choices. The findings of the work suggest that hostilities can drive economic barriers that affect a country's Internet infrastructure.

\vspace{0.5em}
\noindent\textbf{\textit{Network resilience strategy.}} In recent years, several studies have focused on analyzing the resilience of the Interent ecosystem. Focusing on DNS, Allman~\cite{Allman2018} analyzed the structural DNS robustness of the DNS authoritative ecosystem over a period of 9 years and showed the adoption of different resilience techniques by DNS operators. Sommese \etal~\cite{Sommese2021} provided an extensive characterization of the adoption of \textit{Anycast} in DNS authoritative infrastructure and investigated the impact of \textit{Anycast} adoption on the deployment of other resilience techniques.
Finally, Haq \etal~\cite{haq_no_2022} conducted a study on the trend of customer behavior following the 2016 attack on Dyn. Their results confirm that popular domains and specific sectors were more likely to switch away from Dyn after the attack. This paper adopts a similar approach to observe how network managers react to potential disruptions to their networks using Internet measurement data. 

\section{Data Sources}
\label{sec:data_sources}



\noindent\textbf{\textit{Active DNS measurement data.}}
We use DNS measurement data provided by the OpenINTEL project\cite{van_rijswijk-deij_high-performance_2016}.
Using zone files as seeds, it queries selected DNS record types of all domain names under selected TLDs on a daily basis \cite{noauthor_openintel_nodate}\footnote{https://openintel.nl/coverage/}.
In addition, it also queries the infrastructures of domain authoritative name servers, e.g., \texttt{A} records for the name server addresses, daily.
The country-level geolocation data of the infrastructures are inferred from the IP addresses using IP2Location service \cite{ip2locationcom_ip2location_nodate}.

For our analyses, we investigate domain names under new gTLDs within ICANN’s Centralized Zone Data Service (CZDS)\cite{icann_delegated_czds}\footnote{https://newgtlds.icann.org/en/program-status/delegated-strings}.
As of March 2023, there are 1,235 gTLDs registered under the CZDS with $>$20 million unique domain names. 
We use CZDS domains because they offer a more diverse sample of the domain name market. Unlike \texttt{com-net-org} top level domains, CZDS domains serve numerous purposes and represent a wide range of industry sectors, business types, and countries, adding to their value in our analysis.
The dataset contains daily measurements of \texttt{A} records of all apex domains within CZDS for 5 years, starting from January 1st, 2018, to February 28th, 2023.
The starting and end dates mark the 4 years before the conflict started and the first-year mark of the conflict, respectively.
This time-frame was selected to help draw the historical context of the dependency on the network infrastructures in the conflict area long before the conflict and contrast it with the days leading to and during the conflict.

\vspace{0.5em}
\noindent\textbf{\textit{WHOIS data.}}
We use WHOIS records from public RDAP data queried using SpiderRDAP\footnote{https://github.com/gakiwate/SpiderRDAP}--a tool to query RDAP servers at scale \cite{akiwate_spiderrdap_2022}.
However, due to intensive rate limits and to avoid overburdening the RDAP servers, we only scanned a subset ($\sim$100K) of the domains.
We rely on the registrant address field in these data to infer the country of the domain registrant which helps us to understand the digital sovereignty context behind an infrastructure change.

\vspace{-0.3em}
\section{Methodology}
\label{sec:methodology}

\noindent\textbf{\textit{Conflict area.}}
We define \textit{the conflict area} as countries which suffered direct physical impact from the conflict. 
This includes \textit{Russia} and \textit{Ukraine} as the conflicting countries, and the adjacent countries, namely, \textit{Belarus}, \textit{Moldova}, and \textit{Poland}\cite{belarus2022,moldova2022,poland2022}.
We consider physical network infrastructures in these countries possess the high risk of getting destructed by the conflict.

\vspace{0.5em}
\noindent\textbf{\textit{Domain states by infrastructure location.}}
Infrastructure location influences network resilience against physical destruction.
The attacks in the conflict are more likely to destroy infrastructure within the conflict area, rather than outside.
In addition, the presence of redundant network infrastructures, e.g., multiple hosting servers, also ensures higher resilience when an attack destroys the primary infrastructure.
Hence, we define the following states of domain:
\begin{itemize}
    \item \textit{Full inside}: when \textit{all} associated IP addresses of a domain are geo-located to countries \textit{inside} \textit{Conflict area}.
    \item \textit{Partial inside}: when \textit{at least one} associated IP address is geo-located to countries in \textit{Conflict area} and \textit{at least one} outside.
    \item \textit{None inside}: when \textit{all} associated IP addresses are geo-located to countries \textit{outside} \textit{Conflict area}.
\end{itemize}

Another approach to characterizing states of a domain is by looking at the presence of infrastructure particularly in Russia or Ukraine.
The Russian attacks on Ukraine resulted in significant infrastructural damage throughout Ukraine, while the physical impact of Ukrainian attacks was limited to bordering regions of Russia~\cite{noauthor_russian_strikes_2023}.
Therefore, we also check if the domains use infrastructures in Russia, or in Ukraine, or in both countries, or in neither country.

\vspace{0.5em}
\noindent \textbf{\textit{Domain states by infrastructure robustness.}} 
There is another strategy to promote resilience other than moving the physical infrastructure away from the conflict area, namely, to use a more robust infrastructure. 
For instance, network managers could use secondary infrastructures to provide redundancy as a mean for back-up when the primary one suffers a physical damage.
To become even more resilient against physical conflict in a certain area, network managers could use multiple infrastructures located in different countries. 
Finally, network managers could also use network services from 3rd party cloud providers.
To provide a reliable network service for thousands or millions of customers from everywhere, Cloud service providers use a larger scale infrastructure which often consists of a network of edge servers distributed around the globe.

To characterize the robustness of infrastructure that a domain use at a certain time, first we look at two factors: redundancy and global distribution.
Hence, we define the following states of a domain based on its infrastructure robustness:
\begin{itemize}
    \item \textit{Single IP} is when the domain has only 1 unique associated IP address, or \textit{Multiple IP}, when it has multiple unique IP addresses.
    \item \textit{Single Location} is when all IP addresses associated with the domain are geolocated to 1 unique country, or \textit{Multiple Location}, when the IP addresses are geolocated to multiple countries.
\end{itemize}
In addition to that, we also identify the 3rd party DNS services of certain domain names by looking at their authoritative name server addresses at a particular time.

\vspace{0.5em}
\noindent\textbf{\textit{Longitudinal analysis.}}
To give a historical context, we present a longitudinal analysis using the complete dataset of 5 years of measurement in Section~\ref{sec:historical_analysis}.
First, for each daily data, we label all the domains according to the aforementioned classification approaches to characterize the hosting infrastructure.
Then, we count and calculate the proportion of each group relative to the number of unique domains on that day.
Finally, we visualize the time series data with information about the time when the major events happened within the timeline.
In addition to that, we also present a similar analysis of the utilization of 3rd party DNS services. 
We look at the share of six of the largest DNS service providers among CZDS domains and observe its change over time.
However, we do not include the location of DNS infrastructures in our analyses considering the growing implementation of \textit{Anycast}, especially among large DNS service providers, which might introduce noise to our dataset~\cite{Sommese2021}.
Since \textit{Anycast} infrastructures assign the same IP address to multiple servers in different geographical locations, the given geolocation information would be inconsistent depending on the vantage point used.

\vspace{0.5em}
\noindent\textbf{\textit{Temporal comparative analysis.}}
To give a further understanding of how the conflict influences the resilience strategy, we perform an in-depth analysis in Section~\ref{sec:in_depth_analysis} between the daily data of two major events.
The first event is the first Russian military build-up on January 21st, 2021, and the second event is the first-year mark of the conflict, namely, February 24th, 2023.
We chose these two dates to contrast the situations before and after the conflict took place, which we refer to as \textit{initial} and \textit{final} states, respectively.
To account for measurement noise, we use a 5-day window and include all values found in the data two days prior and two days following the date, in addition to the data of that particular date.

We conduct various comparative analyses between the two days. 
\textit{First}, we categorize all the domains and compare their initial and final states based on their infrastructures location.
\textit{Second}, we compare the initial and final locations of the infrastructures among domains with different countries of the registrant.
\textit{Third}, we compare infrastructure robustness between the two days in terms of redundancy and distribution. Table~\ref{tab:data_sources} summarizes the data sources and the data period used for the varying analyses in this paper.

\begin{table}[t]
    \centering
    \resizebox{\linewidth}{!}{
    \begin{tabular}{L{0.5\linewidth}p{0.09\linewidth}p{0.24\linewidth}p{0.26\linewidth}}
    \toprule
    & \textbf{Section} & \textbf{Dataset} & \textbf{Time Period} \\
    \cmidrule{2-4}
    \multicolumn{4}{p{0.4\linewidth}}{\hspace{-0.2cm}\textbf{Longitudinal analyses:}} \\
    \cmidrule{1-1}
    Hosting infrastructure location & \ref{sec:hist_hosting_location} &                     OpenINTEL, IP2Location & Jan'18-Feb'23 \\
    3rd Party Authoritative DNS service & \ref{sec:hist_dns_provider} & OpenINTEL, list of providers' name servers & Jan'18-Feb'23 \\
    \multicolumn{4}{p{0.4\linewidth}}{\hspace{-0.2cm}\textbf{Temporal analyses:}} \\
    \cmidrule{1-1}
    Hosting infrastructure location & \ref{sec:indepth_hosting_location} &  OpenINTEL, IP2Location & Feb'21 \& Feb'23* \\
    Hosting infrastructure location with registrant country &       \ref{sec:indepth_registrant_country} & OpenINTEL, IP2Location, WHOIS & Feb'21 \& Feb'23* \\
    Hosting infrastructure robust\-ness & \ref{sec:indepth_robustness} &  OpenINTEL, IP2Location & Feb'21 \& Feb'23* \\
    \bottomrule
    \multicolumn{4}{p{0.7\linewidth}}{\textit{* using 5-day window, i.e., day$\pm$2}}
    \end{tabular}}
    \caption{Data sources for each analysis.}
    \label{tab:data_sources}
    \vspace{-1em}
\end{table}

\section{Results and Discussion}
\label{sec:result_discussion}

\subsection{Longitudinal analysis}
\label{sec:historical_analysis}

The first objective of this study is to investigate how the dependence on infrastructure in conflict areas changes over the selected period of time and whether recent conflict events coincide with the observed trends. To achieve this, we retrieve daily measurements of \texttt{A} records of all domain names for the entire 5-year time span using a 7-day interval. We classify each domain name according to the approaches presented in Section~\ref{sec:methodology}. Next, we compute the daily number of each domain class (e.g.,\ \textit{Full inside}) and its proportion relative to the total number of unique domain names per day.

\subsubsection{Hosting infrastructure location}
\label{sec:hist_hosting_location}
\begin{figure}[!t]
    \centering
    \includegraphics[width=\linewidth]{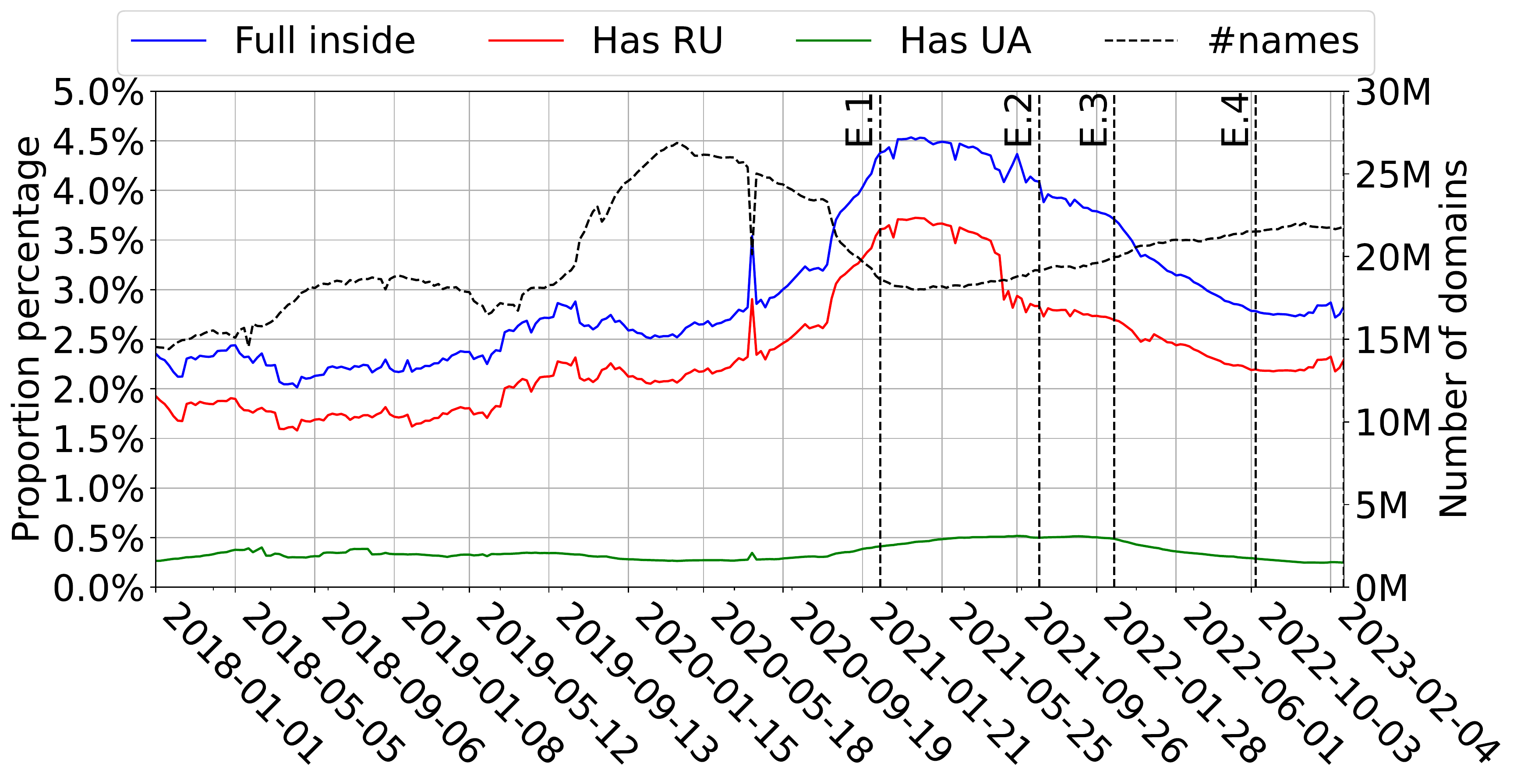}
    \caption{Domain name using hosting infrastructure in the conflict area}
    \label{fig:hist_hosting_zoom_in}
    \vspace{-1em}
\end{figure}

Figure~\ref{fig:hist_hosting_zoom_in} shows the change in the proportion of domain names that fall under some categories according to their hosting infrastructure over time.
The proportion of domain names that use hosting infrastructure outside the conflict area is significantly high, accounting for over 95\% of all domain names. This makes it difficult to observe the other categories when compared with this dominant category. Hence, we do not include these categories (has \textit{None inside} the conflict area and \textit{Has neither} infrastructure in Russia nor Ukraine) in Figure~\ref{fig:hist_hosting_zoom_in} to highlight the changes in the proportion of domain names which use hosting infrastructures inside the conflict area. The vertical dashed lines mark some major events surrounding the Russia-Ukraine conflict, namely:

\begin{itemize}
    \item [E.1.] February 21st, 2021: First Russian military build-up
    \item [E.2.] November 1st, 2021: Second Russian military build-up
    \item [E.3.] February 24th, 2022: Announcement of special military operation (start of the conflict)
    \item [E.4.] October 9th, 2022: First missile attack on Ukrainian residential buildings
\end{itemize}

Observations that we made from Figure~\ref{fig:hist_hosting_zoom_in}:\textit{First,} we can see that the majority of \textit{Full inside} domains use infrastructure from Russia. A small proportion of domains use infrastructure from Ukraine and other countries in the conflict area. \textit{Second,} prior to the first Russian military build-up in February 2021, the proportion of domains that \textit{Full inside} steadily increased. However, not long following this event (E.1.), the proportion of \textit{Full inside} domains consistently decreased until the end of the measurement window which is one year after the conflict started. 
This suggests that the utilization of hosting infrastructures within the conflict area had changed, even prior to the official declaration from President Putin in February 2022, during the initial build-up of tension between the two countries.

\vspace{0.5em}
\noindent\textit{\textbf{Key takeaway:} The percentage of domain names utilizing hosting infrastructures within the conflict area had shown a consistent decrease since the initial Russian military build-up in February 2021, which is approximately one year before the onset of the conflict.}

\begin{figure*}[t]
    \centering
    \includegraphics[width=0.8\linewidth]{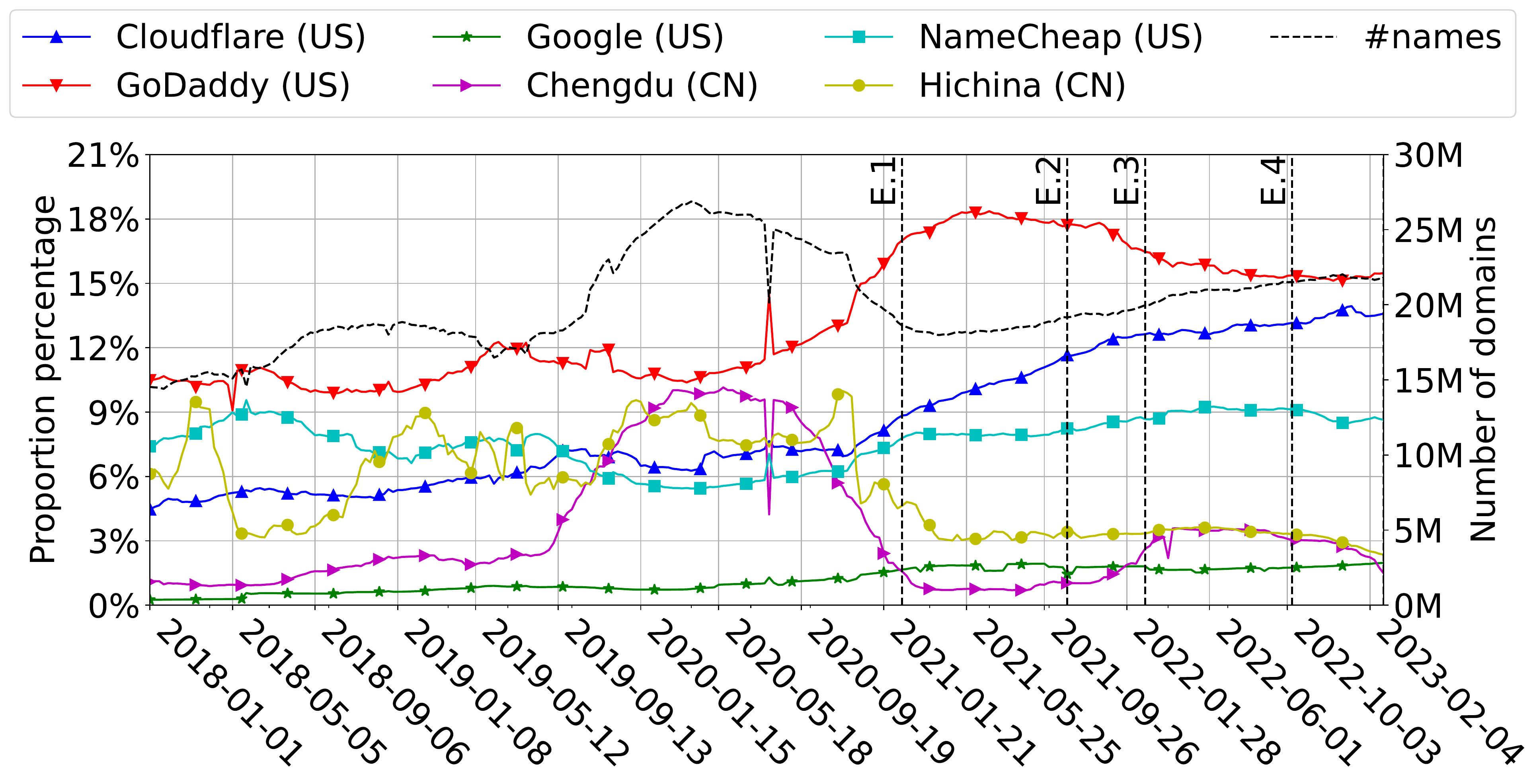}
    \caption{Proportion of customers of six large DNS providers for all domain names. Dashed vertical lines mark the major events.}
    \label{fig:dns_provider_all}
    \vspace{-1em}
\end{figure*}
 
 \subsubsection{3rd Party Authoritative DNS service}
 \label{sec:hist_dns_provider}
Here, we present a longitudinal analysis of the use of 3rd party DNS services (authoritative DNS) among domains that use network infrastructure inside the conflict area.
As noted in the previous section, there appears to be a shift in the trend of using hosting infrastructure within the conflict area, which coincides with the first Russian military build-up near the Ukrainian border.
To promote resilience during the conflict, network managers might also change the infrastructure serving the authoritative DNS for their networks, i.e., by using managed DNS service from 3rd party providers.
Hence, we analyze the use of authoritative DNS services from six large providers over the 5-year time period.
Figure~\ref{fig:dns_provider_all} compares the utilization of managed DNS service providers among all domain names over time.
We infer the DNS providers from the authoritative name server address (\texttt{NS} records) of the domain names using a pre-defined dictionary of name server pattern as shown in the appendix.
Note that one domain might use multiple DNS providers at the same time.
The percentages presented in these figures indicate the proportion of unique domains utilizing these providers relative to the total number of unique CZDS domain names on the given day.


Our observations show that GoDaddy, one of the largest domain registrar companies based in the United States, maintained its position as the largest DNS provider throughout the entire period. However, its market share among CZDS domains began to decrease following the first Russian military build-up in February 2021 (E.1), as depicted in Figure~\ref{fig:dns_provider_all}. In contrast, Cloudflare, one of the largest CDN and DDoS protection service providers from the United States, saw a significant increase in market share compared to the period before the first major event under consideration (E.1). Moreover, the market share of Chengdu, one of the largest ISP and domain registrar companies from China, also started to grow not long after the second Russian military build-up in November 2021 (E.2). The market share of Google (US), NameCheap (US), and Hichina (CN), the other three providers, remained relatively constant and did not show any significant change following any major event in the conflict.


\vspace{0.5em}
\noindent\textit{\textbf{Key takeaway:} We have observed a consistent increase in the adoption of Cloudflare (managed DNS) and an initial increase in the adoption of GoDaddy among the largest authoritative DNS providers. We see a small decrease in the adoption of GoDaddy after second military buildup (E.2.).}



\subsection{Temporal analysis}
\label{sec:in_depth_analysis}
\subsubsection{Hosting infrastructure location}
\label{sec:indepth_hosting_location}

We compare the location of hosting infrastructure between the earliest and the most recent major events to contrast the change before and after the conflict happened which we refer to as the \textit{initial} and the \textit{final} states.
The \texttt{A} records from both days reveal two overlapping sets of domain names indicating some domain names exist on both days and some others do not.
For the former, we classify the domain states on both days using the approach in Section~\ref{sec:methodology} to infer the initial and final states regarding the hosting infrastructure.
We summarized the number of domain names for each possible combination as seen in Table~\ref{tab:hosting_loc_move} together with the distinct number of IP addresses and \texttt{AS}es to provide information on infrastructure concentration.

We made a couple of observations from this analysis.
\textit{First,} out of $\sim$200k existing domains that were initially \textit{Full inside}, $\sim$50k domains or $\sim$25\% \textit{moved away}, i.e., switched to hosting infrastructures outside the conflict area, and $\sim$140k domains or $\sim$70\% \textit{stayed inside}, i.e., continued using infrastructure inside the conflict area.
\textit{Second,} out of $\sim$6.5M existing domains that were initially \textit{None inside}, $\sim$27k domains \textit{moved inside}, i.e., switched to hosting infrastructure inside the conflict area.
It is also worth mentioning that most of the domain names ($\sim$12M) no longer exist and among domain names that still exist, most of them ($\sim$15M) are newly registered, i.e., did not exist in February 2021 (E.1).

One of the strategies to promote network resilience against physical damage in the situation of war is switching to the infrastructure outside the conflict area, as implemented by domain names that \textit{moved away}.
However, this does not explain the domain names that \textit{stayed inside} or even \textit{moved into} the conflict area.
Therefore, we further analyze these actions in the next section by adding the country information of the domain registrant to the dataset to see if these actions might also be influenced by other aspects, for example, protecting the sovereignty of the data and the service against the opposing regimes.

\vspace{0.5em}
\noindent\textit{\textbf{Key takeaway:} Of the approximately 200K existing domains that were initially fully hosted inside the conflict area, around 25\% of them moved their hosting infrastructure away from the conflict area, while the majority remained inside. In addition, around 27K domain names that were initially hosted outside the conflict area moved into the conflict area.}

\begin{table}[!t]
    \centering
    \resizebox{0.95\linewidth}{!}{
    \setlength\tabcolsep{2pt}
    \def\arraystretch{1.2}
    \begin{tabular}{llrrr}
    \toprule
              \textbf{Initial} & \textbf{Final} & 
              \begin{tabular}{rr}\textbf{\# Unique} \\ \textbf{domain} \end{tabular} & 
              \textbf{\# Unique IP} &    
              \textbf{\# Unique AS} \\
    \midrule
    \multirow{4}{*}{Full inside} & Full inside* & \cellcolor{gray!25} 137,576 &    32,559 (23.7\%) &   2,244 (1.6\%) \\
              & Partial inside &             328 &      485 (147.9\%) &    139 (42.4\%) \\
              & None inside** & \cellcolor{gray!25} 49,863 &    19,606 (39.3\%) &   1,003 (2.0\%) \\
              & Not exist &         632,162 &           &        \\
    \cline{1-5}
    \multirow{4}{*}{Partial inside} & Full inside &             336 &       264 (78.6\%) &     80 (23.8\%) \\
              & Partial inside &             149 &      251 (168.5\%) &    106 (71.1\%) \\
              & None inside &             892 &       756 (84.8\%) &    156 (17.5\%) \\
              & Not exist &           9,407 &           &        \\
    \cline{1-5}
    \multirow{4}{*}{None inside} & Full inside*** &  \cellcolor{gray!25} 27,298 &     9,041 (33.1\%) &     780 (2.9\%) \\
              & Partial inside &             581 &      905 (155.8\%) &    284 (48.9\%) \\
              & None inside &       6,432,718 &   907,515 (14.1\%) &  12,630 (0.2\%) \\
              & Not exist &      11,656,987 &           &        \\
    \cline{1-5}
    \multirow{3}{*}{Not exist} & Full inside &         457,563 &    49,509 (10.8\%) &   2,151 (0.5\%) \\
              & Partial inside &           6,669 &    6,797 (101.9\%) &     545 (8.2\%) \\
              & None inside &      15,133,259 &  1,254,706 (8.3\%) &  10,852 (0.1\%) \\
    \bottomrule
    \multicolumn{5}{l}{\textit{*stayed inside the conflict area}} \\
    \multicolumn{5}{l}{\textit{**moved away from the conflict area}} \\ 
    \multicolumn{5}{l}{\textit{***moved in to the conflict area}} \\
    \hline
    \\
    \end{tabular}}
    \caption{Number of domains by initial and final hosting infrastructure locations. The percentages are relative to the number of unique domains per row. Highlighted cells indicate points of observation.}
    \label{tab:hosting_loc_move}
    \vspace{-1em}
\end{table}


\subsubsection{Registrant country}
\label{sec:indepth_registrant_country}
\begin{table}[t!]
    \centering
    \resizebox{0.95\linewidth}{!}{
    \setlength\tabcolsep{3pt}
    \def\arraystretch{1.2}
    \begin{tabular}{ccrrrrrrr}
    \toprule
        \multirow{2}{*}{\begin{tabular}{c}\textbf{Initial}\\\textbf{Location} \end{tabular}} & \multirow{2}{*}{\begin{tabular}{c}\textbf{Registrant}\\ \textbf{Country} \end{tabular}} & \multirow{2}{*}{\textbf{\#names}}& \multicolumn{6}{c}{\textbf{Final Location}} \\
        \cline{4-9}
    
       &  &      &   \textbf{RU} &   \textbf{US} &   \textbf{DE} &   \textbf{UA} &   \textbf{AU} &  \textbf{NL} \\
    \midrule
    \multirow{7}{*}{RU} & RU &  25,249 &  \cellcolor{gray!25} 93\% &   3\% &   1\% &   0\% &   0\% &  2\% \\
       & US &   9,742 & \cellcolor{gray!25} 34\% &  32\% &  16\% &   0\% &  13\% &  2\% \\
       & IS &   1,925 &  14\% &  65\% &   9\% &   1\% &   0\% &  4\% \\
       & JP &   1,511 &   0\% & \cellcolor{gray!25} 24\% & \cellcolor{gray!25} 75\% &   0\% &   0\% &  0\% \\
       & UA &   1,204 & \cellcolor{gray!25} 37\% &  11\% &  11\% &  29\% &   0\% &  5\% \\
       & CN &   1,081 &   7\% & \cellcolor{gray!25} 47\% & \cellcolor{gray!25} 41\% &   0\% &   2\% &  0\% \\
       & PL &     719 &   0\% &   2\% & \cellcolor{gray!25} 82\% &   0\% &   0\% &  0\% \\
    \cline{1-9}
    \multirow{7}{*}{UA} & UA &   4,273 &   1\% &   3\% &   4\% &  86\% &   0\% &  2\% \\
       & US &   1,244 &   5\% &  43\% &  16\% & \cellcolor{gray!25} 15\% &  12\% &  5\% \\
       & IS &     496 &   1\% &  58\% &   7\% &  22\% &   1\% &  6\% \\
       & RU &     471 &  \cellcolor{gray!25} 66\% &  10\% &   5\% &   9\% &   0\% &  5\% \\
       & JP &     240 &   0\% & \cellcolor{gray!25} 18\% & \cellcolor{gray!25} 81\% &   0\% &   0\% &  0\% \\
       & CN &     191 &   0\% & \cellcolor{gray!25} 56\% & \cellcolor{gray!25} 34\% &   1\% &   4\% &  0\% \\
       & PL &      84 &   0\% &   6\% & \cellcolor{gray!25} 65\% &   6\% &   0\% &  1\% \\
    \bottomrule
    \end{tabular}}
    \caption{Changes in hosting infrastructure among domain names which initially fully hosted in Russia or Ukraine characterized by the registrant country. The percentages are relative to the total number of domain names per registrant country initially hosted in either Russia or Ukraine. Highlighted cells indicate points of observation.}
    \label{tab:registrant_move_away}
    \vspace{-1.5em}
\end{table}

To provide context on the identities and affiliations of the domains, we augment the data with information on the countries of the domain registrants. We observe a substantial number of domains that have relocated their hosting infrastructures away from the conflict area. However, we lack an understanding of the underlying reasons that motivated these changes, as well as the reasons for domains that remained in the conflict area. It is possible that network managers made changes to their infrastructure to promote resilience or to protect the sovereignty of digital assets and access. By identifying the country of the domain registrants, we infer their affiliation, thereby enabling us to better understand the motives behind their decisions.

Table~\ref{tab:registrant_move_away} shows the most occuring registrant countries of the domain names that initially used hosting infrastructures in Russia or Ukraine, and compares the initial and final locations of the hosting infrastructures among these domains.
We randomly sampled $\sim$100k out of $\sim$200k domains that were using hosting infrastructures inside the conflict area and then extracted their registrant country from \texttt{WHOIS} records with the help of SpiderRDAP.
As mentioned in Section~\ref{sec:methodology}, we intentionally avoided overburdening the public RDAP servers by not querying all the domains.
Out of $\sim$100k sampled domains, there are $\sim$24k domains ($\sim$27\%) excluded from this analysis due to the absence of country information in their \texttt{WHOIS} registrant address fields which are mostly redacted due to privacy concerns.
Finally, we used geolocations of their hosting IP addresses to infer the initial and the final infrastructure locations.
We use the registrant country to associate the ownership of a domain name, for example, we refer to \textit{Russian domains} as the domains whose registrants are from Russia.

\newpage
Our observations from Table~\ref{tab:registrant_move_away} are: \textit{First,} the majority of Russian domains which initially used infrastructure in Ukraine switched to hosting infrastructure in Russia (66\%), and those which already used Russian infrastructure continued to do so (93\%).
\textit{Second,} only 37\% of the Ukrainian domains which initially hosted in Russia stayed, while the other 63\% moved away to infrastructures from other countries.
The actions to use local hosting infrastructures and to avoid using infrastructure from the opposing country are most probably driven by the need to maintain sovereignty, i.e., full control of their own network, in order to avoid state-sponsored cyberattacks, such as snooping confidential data or geoblocking \cite{lin2012cyber}.
However, the actions of Ukrainian domains that \textit{moved away} from Russia to countries outside the conflict area might also be motivated by a resilience strategy to avoid the impact of the physical conflict.
Nevertheless, it is also possible to still use local but more resilient infrastructure, for example, by using infrastructure which geographically far from the conflict epicenter or by using redundant infrastructures.

We also made other interesting observations regarding domains with other registrant countries besides Russia and Ukraine.
\textit{First,} Japanese domains changed in a similar pattern regardless of their initial infrastructure locations. They moved away almost exclusively to two specific countries outside the conflict area, namely, Germany and the United States. 
Further analysis is required to confirm if this change is a state-instructed action, e.g., to promote resilience or sovereignty, complied by almost all Japanese domains.
\begin{figure}[t]
    \centering
    \subfloat[Move away]{\includegraphics[width=0.85\linewidth]{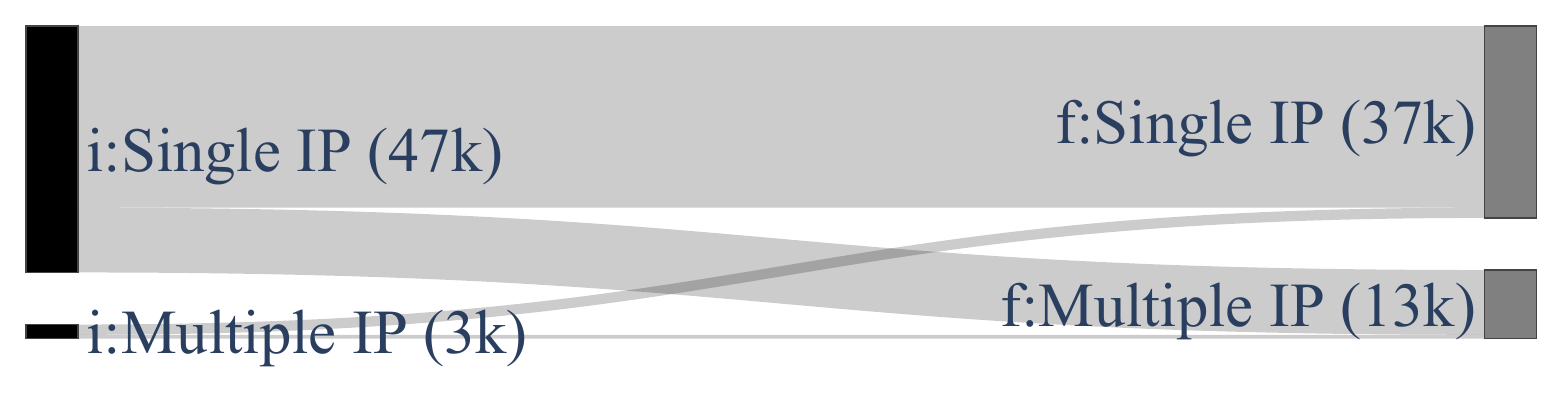}%
    \label{fig:sankey_move_away_ip_duplication}}
    \hfil
    \subfloat[Stay inside or move into]{\includegraphics[width=0.85\linewidth]{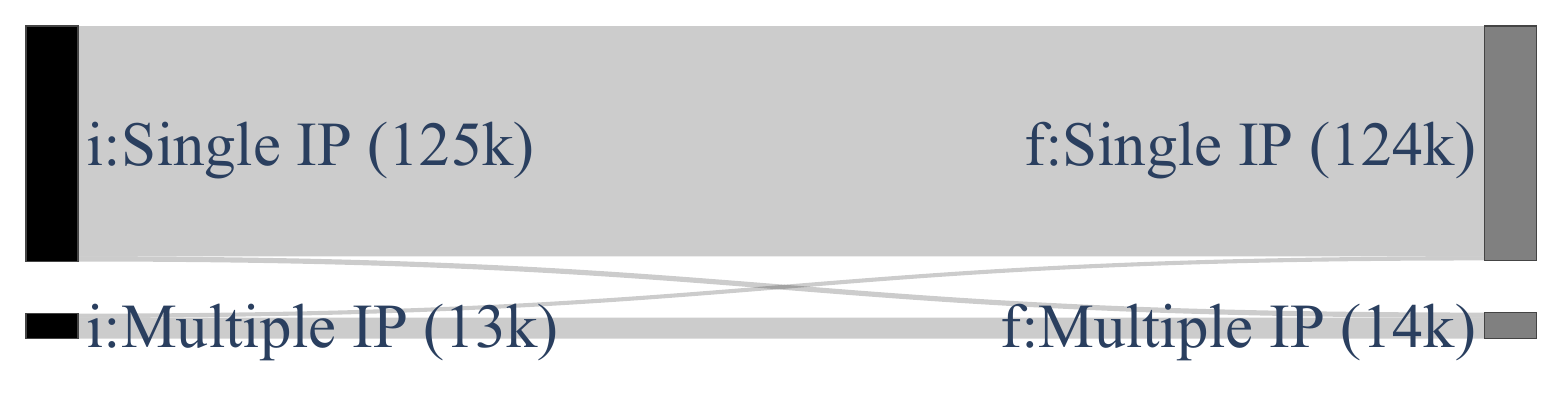}%
    \label{fig:sankey_move_in_stay_ip_duplication}}
    \caption{Changes in hosting infrastructure \textit{redundancy} among domains that \textit{(a) move away from} and \textit{(b) stay inside or move into} the conflict area. \textit{i} and \textit{f} indicate \textit{initial} and \textit{final} conditions, respectively.}
    \label{fig:sankey_hosting_duplication}
    \vspace{-1em}
\end{figure}
A similar pattern is also found among Chinese domains, namely, the majority of Chinese domains, regardless of their initial location, moved away from the conflict area to either Germany or the United States.
\textit{Second,} most of the Polish domains moved away from the conflict area to Germany.
\textit{Finally,} only a small portion of American domains stayed in Russia (34\%) and Ukraine (15\%) while the rest moved away from the conflict area. 
It is also worth mentioning that NameCheap, one of the largest domain registrars based in the US, uses a domain privacy service that acts as a proxy for the actual domain registrant and redacts the registrant information with the contact information of WithheldforPrivacy\footnote{https://withheldforprivacy.com/} -- the NameCheap owned privacy provider based in Iceland. This partially explains the large portion of Icelandic domains that move to the US.

\vspace{0.5em}
\noindent\textit{\textbf{Key takeaway:} The majority of Russian domains that were initially hosted in Ukraine have moved back to Russia, whereas Ukrainian domains that were hosted in Russia have moved away from Russia to either return to Ukraine or to relocate to other countries outside the conflict area. Additionally, most domains of other countries that were hosted in either Russia or Ukraine have also moved away from these countries, but with varying location preferences.}
\vspace{0.5em}





\subsubsection{Network infrastructure robustness}
\label{sec:indepth_robustness}

In this section, we present our analysis of the network infrastructure change regarding the presence of redundancy and the physical distribution of the infrastructure.
In the previous sections, we discussed that some domain names moved away to hosting infrastructure outside the conflict area after the conflict but some other domains did not.
However, the latter decision does not always mean that the network managers of these domains did not take any measures to promote resilience in the situation of war.
Therefore, we also analyze the change in their infrastructure redundancy and distribution after the conflict in the following section.
%

\vspace{0.5em}
\noindent\textbf{\textit{Network infrastructure redundancy.}}
We distinguish between domain names that moved their hosting infrastructure away from the conflict area and those that stayed or moved into the conflict area because the former took an arguably more resilient choice than the latter.
Figure~\ref{fig:sankey_hosting_duplication} shows how infrastructure redundancy of domains in either category change after the conflict. 
We may observe that a significant number ($\sim$13k) of domains that moved their hosting infrastructure away also adopted redundant hosting infrastructure.
Meanwhile, similar action is not observed among domains that are still using hosting infrastructures from inside the conflict area.

\begin{figure}[t]
    \centering
    \subfloat[Move away]{\includegraphics[width=0.85\linewidth]{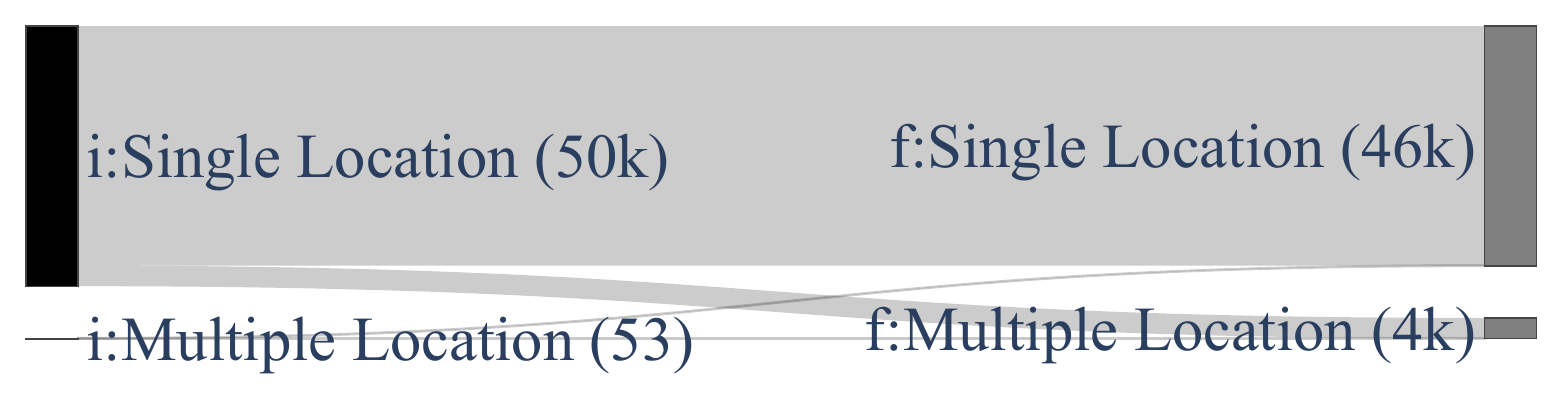}%
    \label{fig:sankey_move_away_ip_distribution}}
    \hfil
    \subfloat[Stay inside or move into]{\includegraphics[width=0.85\linewidth]{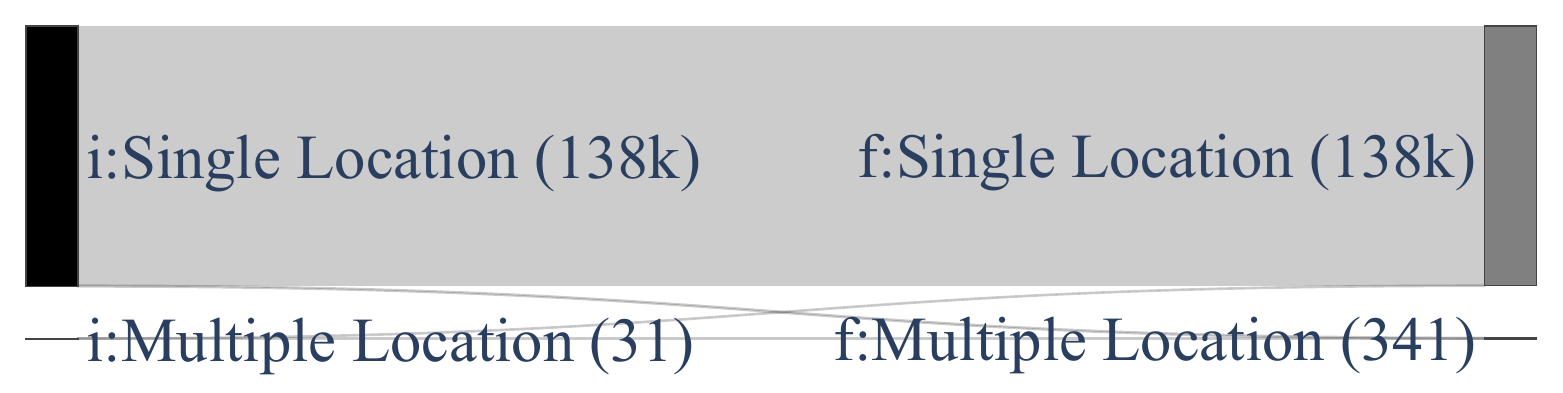}%
    \label{fig:sankey_move_in_stay_ip_distribution}}
    \caption{Changes in hosting infrastructure \textit{distribution} among domains that \textit{(a) move away from} and \textit{(b) stay inside or move into} the conflict area. \textit{i} and \textit{f} indicate \textit{initial} and \textit{final} conditions, respectively.}
    \label{fig:sankey_hosting_distribution}
    \vspace{-1em}
\end{figure}

\vspace{0.4em}
\noindent\textbf{\textit{Network infrastructure distribution.}}
We do the same analysis for hosting infrastructure distribution which is summarized in Figure~\ref{fig:sankey_hosting_distribution}.
Regarding infrastructure distribution, we also observe a similar pattern with infrastructure redundancy, namely, a significant number of domains ($\sim$4k) that moved away also adopted redundant hosting infrastructure \textit{from different countries}, while similar action is not observed among domains that still using infrastructures from inside the conflict area.

\vspace{0.5em}
\noindent\textit{\textbf{Key takeaway:} Domains that adopted redundant and geographically distributed hosting infrastructures are mostly domains that also moved their infrastructure away.}

\subsection{Conclusion}

In this paper, we analyze DNS measurement data over time to study the impact of the Russia-Ukrain conflict on network infrastructures supporting ICANN's CZDS new gTLDs. Our 5-year longitudinal analysis shows a decrease in domain names hosted within the conflict area, starting before the recent conflict began. We observe a shift in user proportions among DNS service providers after the Russian military build-ups in February and November 2021. Comparing data from one year before and after the conflict, we find that 25\% of approximately 200K domains initially hosted in the conflict area moved their hosting location elsewhere, while some 27K domains moved their hosting infrastructure into the conflict area. Our analysis also reveals changes in domain hosting locations based on registrant countries, with domains with Russian registrants moving back to Russia and domains with Ukrainian registrants moving away from Russian infrastructure. Domains with registrants from other countries initially hosted in conflict are also moved away. Our findings suggest that the conflict prompted network operators to take proactive measures to ensure resilience and in some cases protect their networks' sovereignty.

\vspace{0.5em}
\noindent\textbf{Limitations:} Acknowledging the limitations of our study, it is important to note that we do not differentiate between private and public shared hosting infrastructures. As a result, we do not attribute the observed changes in network infrastructure to individual domain managers or third-party hosting providers.

\vspace{0.5em}
\noindent\textbf{Future works:} 
To further understand the decisions made by the network managers, we plan to integrate additional information reflecting other characteristics such as industry sectors, maliciousness, and popularity of the domain names as performed in~\cite{haq_no_2022}.
In addition, we also plan to enrich our dataset with data regarding service blocking of the domain names over time using Internet censorship data e.g., OONI dataset~\cite{about_ooni}\footnote{https://ooni.org/about/}. 
Using this information, we aim to identify domain movements preceded by a service disruption e.g., state-sponsored geo-blocking, to distinguish between the reactive and the proactive measures of the network managers.




\section*{Acknowledgements}
This work is part of the NWO: MASCOT project, which is funded by the Netherlands Organization for Scientific Research (CS.014).
This research was made possible by OpenINTEL, a joint project of the University of Twente, SURF, SIDN, and NLnet Labs.

\bibliographystyle{IEEEtran}
\bibliography{references.bib}

\begin{thebibliography}{10}
\providecommand{\url}[1]{#1}
\csname url@samestyle\endcsname
\providecommand{\newblock}{\relax}
\providecommand{\bibinfo}[2]{#2}
\providecommand{\BIBentrySTDinterwordspacing}{\spaceskip=0pt\relax}
\providecommand{\BIBentryALTinterwordstretchfactor}{4}
\providecommand{\BIBentryALTinterwordspacing}{\spaceskip=\fontdimen2\font plus
\BIBentryALTinterwordstretchfactor\fontdimen3\font minus
  \fontdimen4\font\relax}
\providecommand{\BIBforeignlanguage}[2]{{%
\expandafter\ifx\csname l@#1\endcsname\relax
\typeout{** WARNING: IEEEtran.bst: No hyphenation pattern has been}%
\typeout{** loaded for the language `#1'. Using the pattern for}%
\typeout{** the default language instead.}%
\else
\language=\csname l@#1\endcsname
\fi
#2}}
\providecommand{\BIBdecl}{\relax}
\BIBdecl

\bibitem{lin2012cyber}
H.~Lin, ``Cyber conflict and international humanitarian law,''
  \emph{International review of the Red Cross}, vol.~94, no. 886, pp. 515--531,
  2012.

\bibitem{bergengruen_battle_2022}
\BIBentryALTinterwordspacing
V.~Bergengruen, ``\BIBforeignlanguage{en}{The {Battle} for {Control} {Over}
  {Ukraine}'s {Internet}},'' \emph{\BIBforeignlanguage{en}{Time}}, Oct. 2022.
  [Online]. Available:
  \url{https://time.com/6222111/ukraine-internet-russia-reclaimed-territory/}
\BIBentrySTDinterwordspacing

\bibitem{noauthor_internet_nodate}
\BIBentryALTinterwordspacing
``Internet {Resilience}.'' [Online]. Available:
  \url{https://pulse.internetsociety.org/resilience}
\BIBentrySTDinterwordspacing

\bibitem{50ab3fadbc7341e9bf8a5c757b757fd1}
A.~Abhishta, R.~Van Rijswijk-Deij, and L.~J. Nieuwenhuis,
  ``\BIBforeignlanguage{English}{Measuring the impact of a successful ddos
  attack on the customer behaviour of managed dns service providers},''
  \emph{\BIBforeignlanguage{English}{Computer communication review}}, vol.~48,
  no.~5, pp. 70--76, October 2018.

\bibitem{noauthor_donbas_war_2023}
\BIBentryALTinterwordspacing
``\BIBforeignlanguage{en}{War in {Donbas} (2014–2022)},'' May 2023, page
  Version ID: 1154048331. [Online]. Available:
  \url{https://en.wikipedia.org/w/index.php?title=War_in_Donbas_(2014%E2%80%932022)&oldid=1154048331}
\BIBentrySTDinterwordspacing

\bibitem{noauthor_untold_2022}
\BIBentryALTinterwordspacing
``\BIBforeignlanguage{en}{The {Untold} {Story} of the {Ukraine} {Crisis}},''
  Feb. 2022. [Online]. Available:
  \url{https://time.com/6144109/russia-ukraine-vladimir-putin-viktor-medvedchuk/}
\BIBentrySTDinterwordspacing

\bibitem{noauthor_prelude_2023}
\BIBentryALTinterwordspacing
``\BIBforeignlanguage{en}{Prelude to the {Russian} invasion of {Ukraine}},''
  May 2023, page Version ID: 1154104557. [Online]. Available:
  \url{https://en.wikipedia.org/w/index.php?title=Prelude_to_the_Russian_invasion_of_Ukraine&oldid=1154104557}
\BIBentrySTDinterwordspacing

\bibitem{noauthor_putin_announcement_2023}
\BIBentryALTinterwordspacing
``\BIBforeignlanguage{en}{On conducting a special military operation},'' May
  2023, page Version ID: 1153623043. [Online]. Available:
  \url{https://en.wikipedia.org/w/index.php?title=On_conducting_a_special_military_operation&oldid=1153623043}
\BIBentrySTDinterwordspacing

\bibitem{noauthor_russian_strikes_2023}
\BIBentryALTinterwordspacing
``\BIBforeignlanguage{en}{2022–2023 {Russian} strikes against {Ukrainian}
  infrastructure},'' May 2023, page Version ID: 1154084025. [Online].
  Available:
  \url{https://en.wikipedia.org/w/index.php?title=2022%E2%80%932023_Russian_strikes_against_Ukrainian_infrastructure&oldid=1154084025}
\BIBentrySTDinterwordspacing

\bibitem{noauthor_russia_energy_attack_2022}
\BIBentryALTinterwordspacing
``\BIBforeignlanguage{en}{Russia unleashes biggest attacks in {Ukraine} in
  months},'' Oct. 2022, section: Russia-Ukraine war. [Online]. Available:
  \url{https://apnews.com/article/russia-ukraine-kyiv-government-and-politics-8f625861590b9e0dd336dabc0880ac8c}
\BIBentrySTDinterwordspacing

\bibitem{noauthor_ukraine_nodate}
\BIBentryALTinterwordspacing
``\BIBforeignlanguage{en-US}{Ukraine {Data} {Centers} {Became} {Physical}
  {Targets} {When} {Cyberattacks} {Failed}}.'' [Online]. Available:
  \url{https://www.meritalk.com/articles/ukraine-data-centers-became-physical-targets-when-cyber-attacks-failed/}
\BIBentrySTDinterwordspacing

\bibitem{sanger_as_2022}
\BIBentryALTinterwordspacing
D.~E. Sanger, J.~E. Barnes, and K.~Conger, ``\BIBforeignlanguage{en-US}{As
  {Tanks} {Rolled} {Into} {Ukraine}, {So} {Did} {Malware}. {Then} {Microsoft}
  {Entered} the {War}.}'' \emph{\BIBforeignlanguage{en-US}{The New York
  Times}}, Mar. 2022. [Online]. Available:
  \url{https://www.nytimes.com/2022/02/28/us/politics/ukraine-russia-microsoft.html}
\BIBentrySTDinterwordspacing

\bibitem{stupp_ukraine_nodate}
\BIBentryALTinterwordspacing
C.~Stupp, ``\BIBforeignlanguage{en-US}{Ukraine {Has} {Begun} {Moving}
  {Sensitive} {Data} {Outside} {Its} {Borders}},'' section: WSJ Pro. [Online].
  Available:
  \url{https://www.wsj.com/articles/ukraine-has-begun-moving-sensitive-data-outside-its-borders-11655199002}
\BIBentrySTDinterwordspacing

\bibitem{jonker_where_2022}
\BIBentryALTinterwordspacing
M.~Jonker, G.~Akiwate, A.~Affinito, k.~Claffy, A.~Botta, G.~M. Voelker, R.~van
  Rijswijk-Deij, and S.~Savage, ``Where .ru? assessing the impact of conflict
  on russian domain infrastructure,'' in \emph{Proceedings of the 22nd {ACM}
  {Internet} {Measurement} {Conference}}, ser. {IMC} '22.\hskip 1em plus 0.5em
  minus 0.4em\relax New York, NY, USA: Association for Computing Machinery,
  Oct. 2022, pp. 159--165. [Online]. Available:
  \url{https://dl.acm.org/doi/10.1145/3517745.3561423}
\BIBentrySTDinterwordspacing

\bibitem{Allman2018}
M.~Allman, ``{Comments on {{DNS}} Robustness},'' in \emph{Proceedings of the
  Internet Measurement Conference}, ser. IMC '18, 2018.

\bibitem{Sommese2021}
R.~Sommese, G.~Akiwate, M.~Jonker, G.~Moura, M.~Davids, R.~{van Rijswijk -
  Deij}, G.~Voelker, S.~Savage, K.~Claffy, and A.~Sperotto,
  ``\BIBforeignlanguage{English}{{Characterization of Anycast Adoption in the
  DNS Authoritative Infrastructure}},'' in
  \emph{\BIBforeignlanguage{English}{5th Network Traffic Measurement and
  Analysis Conference, TMA 2021}}.\hskip 1em plus 0.5em minus 0.4em\relax IFIP,
  2021.

\bibitem{haq_no_2022}
M.~Y.~M. Haq, M.~Jonker, R.~Van Rijswijk-Deij, K.~Claffy, L.~J. Nieuwenhuis,
  and A.~Abhishta, ``No {Time} for {Downtime}: {Understanding} {Post}-{Attack}
  {Behaviors} by {Customers} of {Managed} {DNS} {Providers},'' in \emph{2022
  {IEEE} {European} {Symposium} on {Security} and {Privacy} {Workshops}
  ({EuroS}\&{PW})}, Jun. 2022, pp. 322--331, iSSN: 2768-0657.

\bibitem{van_rijswijk-deij_high-performance_2016}
R.~van Rijswijk-Deij, M.~Jonker, A.~Sperotto, and A.~Pras, ``A
  {High}-{Performance}, {Scalable} {Infrastructure} for {Large}-{Scale}
  {Active} {DNS} {Measurements},'' \emph{IEEE Journal on Selected Areas in
  Communications}, vol.~34, no.~6, pp. 1877--1888, Jun. 2016, conference Name:
  IEEE Journal on Selected Areas in Communications.

\bibitem{noauthor_openintel_nodate}
\BIBentryALTinterwordspacing
``\BIBforeignlanguage{en}{{OpenINTEL}: {Active} {DNS} {Measurement}
  {Project}}.'' [Online]. Available: \url{https://openintel.nl/coverage/}
\BIBentrySTDinterwordspacing

\bibitem{ip2locationcom_ip2location_nodate}
\BIBentryALTinterwordspacing
IP2Location.com, ``\BIBforeignlanguage{en}{{IP2Location}™ {IP} {Address}
  {GeoLocation} {Database}}.'' [Online]. Available:
  \url{https://www.ip2location.com/database/ip2location/}
\BIBentrySTDinterwordspacing

\bibitem{icann_delegated_czds}
\BIBentryALTinterwordspacing
``Delegated {Strings} {\textbar} {ICANN} {New} {gTLDs}.'' [Online]. Available:
  \url{https://newgtlds.icann.org/en/program-status/delegated-strings}
\BIBentrySTDinterwordspacing

\bibitem{akiwate_spiderrdap_2022}
\BIBentryALTinterwordspacing
G.~Akiwate, ``{SpiderRDAP},'' Jun. 2022, original-date: 2019-10-13T05:05:16Z.
  [Online]. Available: \url{https://github.com/gakiwate/SpiderRDAP}
\BIBentrySTDinterwordspacing

\bibitem{belarus2022}
\BIBentryALTinterwordspacing
``Belarus protests to {Ukraine} after downing stray air defence missile
  {\textbar} {Reuters},'' Dec. 2022. [Online]. Available:
  \url{https://web.archive.org/web/20221231025540/https://www.reuters.com/world/europe/ukrainian-air-defence-missile-lands-belarus-belta-2022-12-29/}
\BIBentrySTDinterwordspacing

\bibitem{moldova2022}
\BIBentryALTinterwordspacing
``O rachetă rusească a căzut pe teritoriul {R}. {Moldova}. {Mai} multe case
  din {Naslavcea}, avariate,'' Dec. 2022. [Online]. Available:
  \url{https://web.archive.org/web/20221206094532/https://moldova.europalibera.org/a/explozii-la-frontiera-de-nord-cu-ucraina-mai-multe-case-din-naslavcea-avariate/32108598.html}
\BIBentrySTDinterwordspacing

\bibitem{poland2022}
\BIBentryALTinterwordspacing
``Russian missile hits {Nato} member {Poland}, leaving two dead.'' [Online].
  Available:
  \url{https://web.archive.org/web/20221116212209/https://www.telegraph.co.uk/world-news/2022/11/15/two-russian-rockets-hit-poland-killing-two/}
\BIBentrySTDinterwordspacing

\bibitem{about_ooni}
\BIBentryALTinterwordspacing
``\BIBforeignlanguage{en}{About {OONI}}.'' [Online]. Available:
  \url{https://ooni.org/about/}
\BIBentrySTDinterwordspacing

\end{thebibliography}
\appendices
\section{Name server patterns}
We use the patterns shown in Table \ref{tab:ns_pattern} below to infer the DNS providers from the domains \texttt{NS} addresses.
\begin{table}[h]
    \centering
    \setlength\tabcolsep{3pt}
    \def\arraystretch{1.2}
    \begin{tabular}{ll}
    \toprule
             \textbf{Provider} &             \textbf{Name server pattern} \\
    \midrule
      Cloudflare (US) &         \texttt{*.cloudflare.*} \\
         GoDaddy (US) &      \texttt{*.domaincontrol.*} \\
          Google (US) &      \texttt{*.googledomains.*} \\
         Chengdu (CN) &        \texttt{*.myhostadmin.*} \\
       NameCheap (US) &  \texttt{*.registrar-servers.*} \\
         Hichina (CN) &            \texttt{*.hichina.*} \\
    \bottomrule
    \end{tabular}
    \caption{Name server address pattern for selected DNS service providers.}
    \label{tab:ns_pattern}
\end{table}

\end{document}